\documentclass[copyright]{eptcs}
\providecommand{\event}{MBT Workshop} 
\usepackage{graphicx}
\usepackage{breakurl}             
\usepackage{enumitem}
\usepackage{flushend}
\usepackage{booktabs}
\usepackage{caption}

\title{Potential Errors and Test Assessment in\\ Software Product Line Engineering}
\author{Hartmut Lackner
\institute{Humboldt-Universit{\"{a}}t zu Berlin\\
Germany}
\email{lackner@informatik.hu-berlin.de}
\and
Martin Schmidt
\institute{Humboldt-Universit{\"{a}}t zu Berlin\\
Germany}
\email{schmidma@informatik.hu-berlin.de}
}
\def\titlerunning{Potential Errors and Test Assessment in Software Product Line Engineering}
\def\authorrunning{Hartmut Lackner \& Martin Schmidt}
\begin{document}
\maketitle

\begin{abstract}

Software product lines (SPL) are a method for the development of variant-rich software systems. 
Compared to non-variable systems, testing SPLs is extensive due to an increasingly amount of possible products.
Different approaches exist for testing SPLs, but there is less research for assessing the quality of these tests by means of error detection capability.
Such test assessment is based on error injection into correct version of the system under test.
However to our knowledge, potential errors in SPL engineering have never been systematically identified before.

This article presents an overview over existing paradigms for specifying software product lines and the errors that can occur during the respective specification processes.
For assessment of test quality, we leverage mutation testing techniques to SPL engineering and implement the identified errors as mutation operators.
This allows us to run existing tests against defective products for the purpose of test assessment.
From the results, we draw conclusions about the error-proneness of the surveyed SPL design paradigms and how quality of SPL tests can be improved.

\end{abstract}

\section{Introduction}

Software product line (SPL) engineering is an emerging method for the development of variant-rich software systems. 
Based on a SPL specification single products can be configured and derived.
SPL engineering is a systematic and planned process to reuse software artifacts most efficiently~\cite{Clements.2009}. 
This also includes quality assurance, where one of the most important ones is testing.
But testing a SPL is different from testing non-variable systems and thus is investigated intensively~\cite{Olimpiew.2005, Engstrom.2011c, Oster.2011, Lochau.2012}.
Challenges in testing SPLs are the selection of products for testing and the design of tests from the SPL's specification.

Though there are many methods proposed for testing a product line, until now, quality assessment of tests was limited to mutating individual products of the SPL.
This approach has two major drawbacks: first, developers can introduce errors on all kinds of artifacts, not only on final products specifications. For better understanding we analyze different design paradigms for the specification of products from a SPL and the errors that can occur during the respective design processes. From the results, we developed mutation operators for variability models, and domain models that mimic possible faults in these models.

Secondly, the selection of products and subsequently its mutations is biased by the available tests, since only products for which tests are available will be tested.
Therefore, mutation analysis assesses the quality of the tests for particular products, but not for the whole SPL.
In contrast, we define a mutation system and operators on the domain engineering-level.
This enables us to assess the test quality independently from the tested products. 
Subsequently, the test quality over the complete SPL is assessed.

\begin{figure}
\centering
\begin{minipage}[b]{.5\textwidth}
\centering
\includegraphics[width=0.9\textwidth]{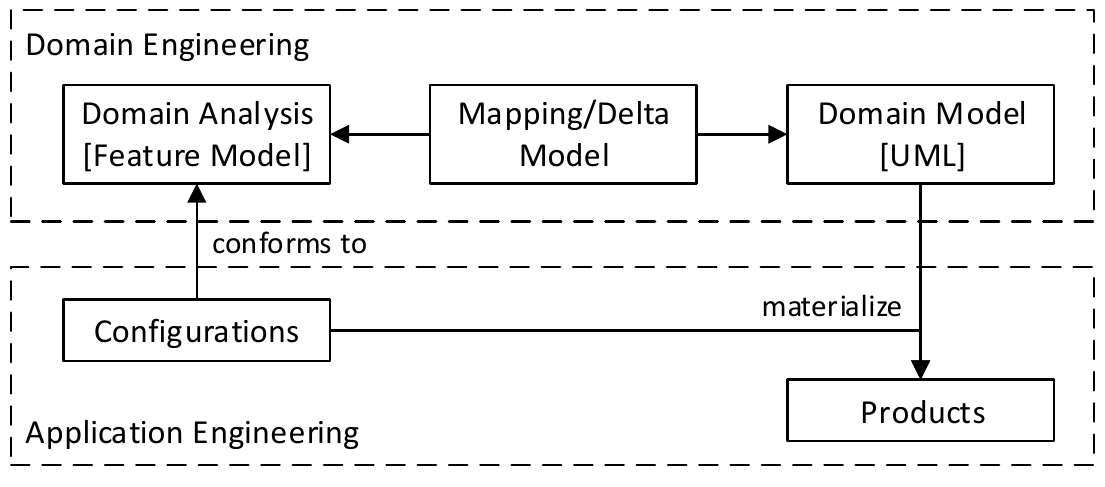}
\caption{Process of SPL engineering}
\label{fig:foda}
\end{minipage}%
\begin{minipage}[b]{.5\textwidth}
\centering
\includegraphics[width=0.9\linewidth]{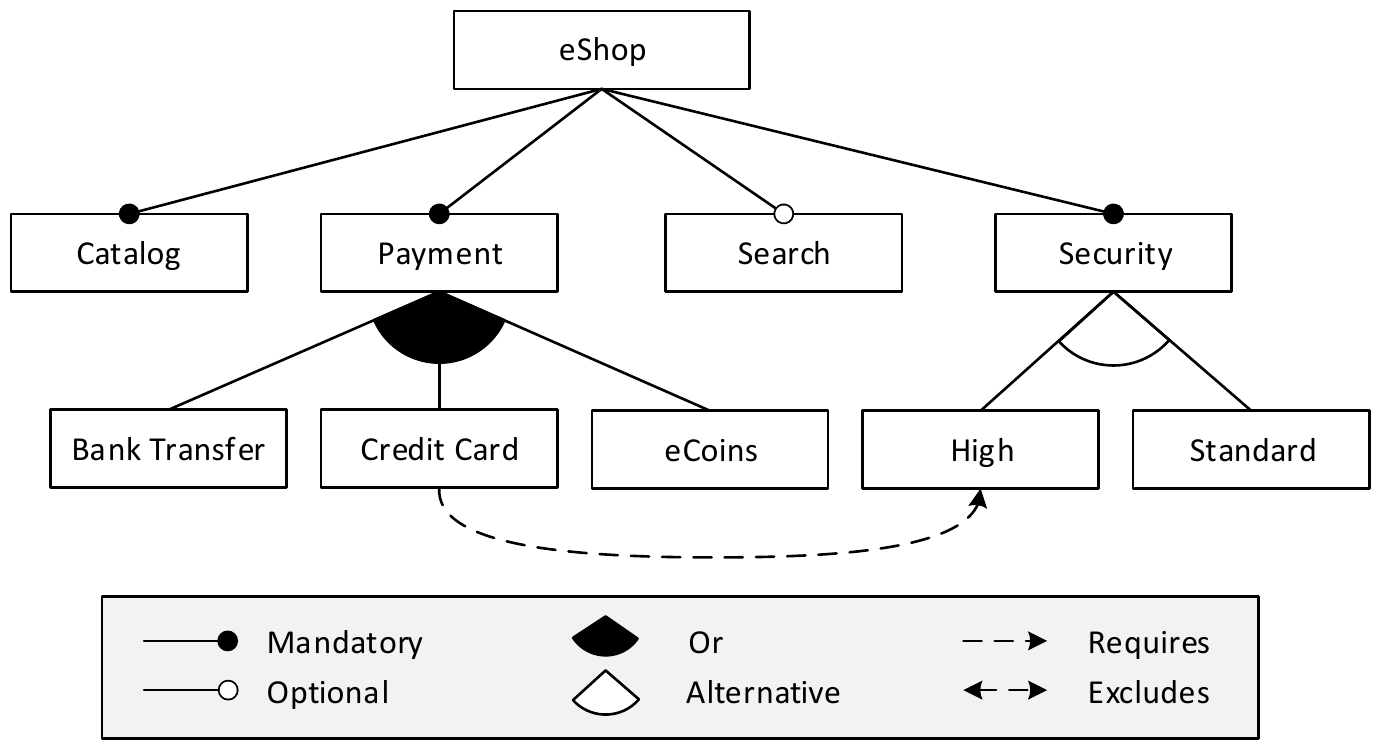}
\captionof{figure}{A feature model for the eShop example.}
\label{fig:featuremodel}
\end{minipage}%
\end{figure}

The remainder of this article is structured as follows: In Section~\ref{sec.preliminaries} we summarize the foundations of SPL engineering and test assessment.
In Sec.~\ref{sec.potentialerrors} we define and classify kinds of errors. 
We present our SPL test assessment system and the evaluation of three examples in Sec.~\ref{sec.productlinetestassessment}. 
Eventually, we show related work in Sec.~\ref{sec.relatedwork} and conclude in Sec.~\ref{sec.conclusions}.


\section{Preliminaries}
\label{sec.preliminaries}

In this section, we present the foundations that our work is based on. First, we give a short introduction for model-based product line engineering. The second part is about mutation analysis. The third part deals with potential errors and mutation in software development.

\subsection{Model-based Product Line Engineering}

Individual customer expectations and the reuse of existing assets in a product's design are two driving factors for the emergence of product line engineering: increasing the number of product features while keeping system engineering costs at a reasonable level. 
In terms of software engineering, a SPL is a set of related software products that share a common core of software assets (commonalities), but can be distinguished (variabilities)~\cite{Pohl.2005}.

The definition and realization of commonalities and variabilities is the process of domain engineering.
Actual products are built during application engineering (cf. fig.~\ref{fig:foda}).
Here, products are built by reusing domain artifacts and exploiting the product line variability.

Like many methodologies, SPL engineering can be supported by model-based abstractions such as feature models. 
Feature models offer a way to overcome the aforementioned challenges by facilitating the explicit design of global system variation points~\cite{Kang.1990}. In consequence, variation points are not spread across one or multiple domain models anymore, but instead linked to one core of variability description.

A feature model has a tree structure in which a feature can be decomposed into sub-features. 
Fig.~\ref{fig:featuremodel} shows an example feature model, that will also be used as an example in section \ref{sec.productlinetestassessment}. 
A parent feature can have the following relations to its sub-features: (a)~\emph{Mandatory}: child feature is required, (b)~\emph{Optional}: child feature is optional, (c)~\emph{Or}: at least one of the children features must be selected, and (d)~\emph{Alternative}: exactly one of the children features must be selected. Furthermore, one may specify additional (cross-tree) constraints between two features A and B: (i)~A \emph{requires} B: the selection of A implies the selection of B, and (ii)~A \emph{excludes} B: both features A and B must not be selected for the same product.

\begin{figure}
\centering
\begin{minipage}[t]{.5\textwidth}
\centering
\includegraphics[width=0.9\textwidth]{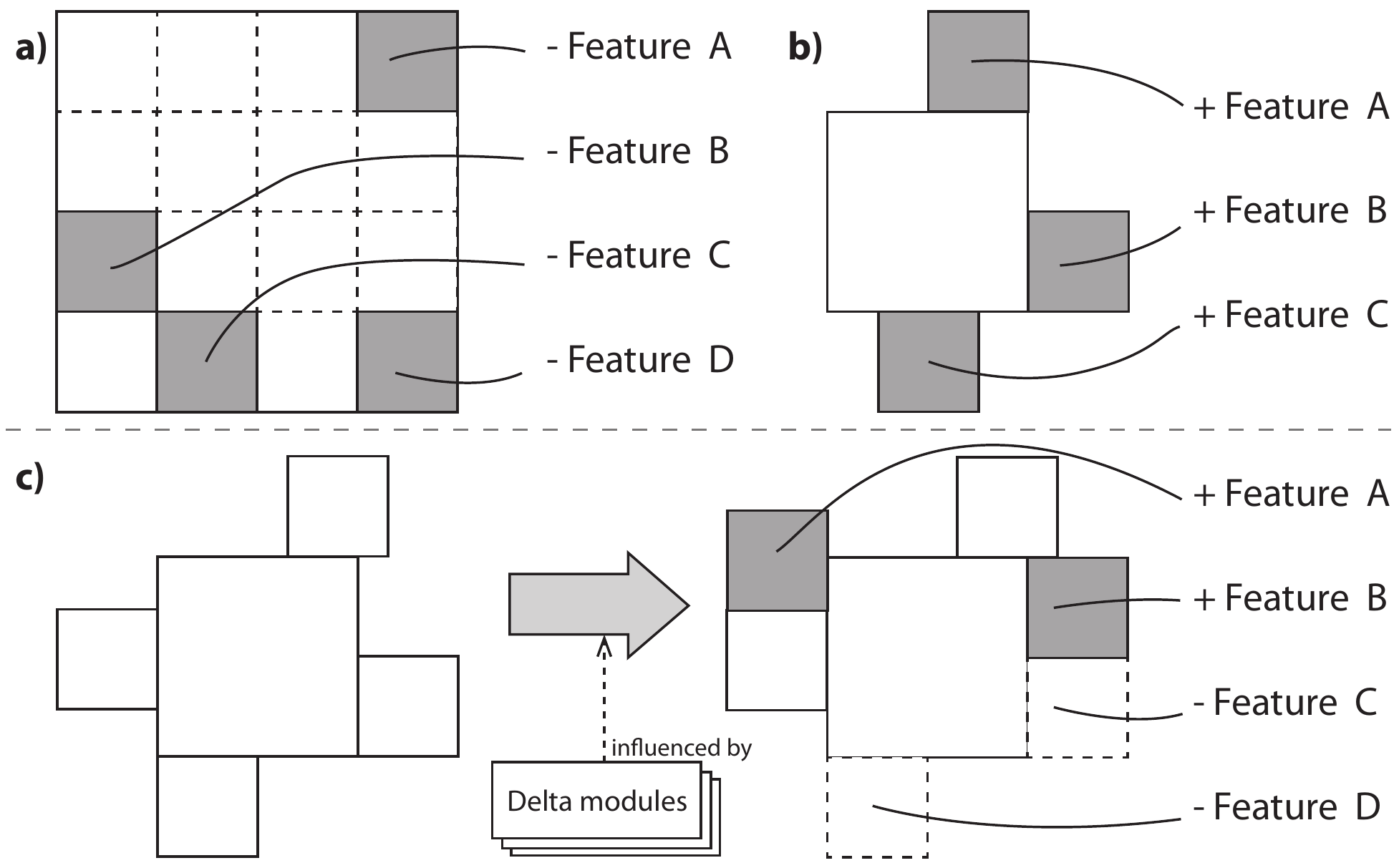}
\caption{Negative (a), positive (b) variability (based on~\cite{Groher.2007}) and delta modeling (c)}
\label{fig:variabilitydelta}
\end{minipage}%
\begin{minipage}[t]{.5\textwidth}
\centering
\includegraphics[width=0.9\linewidth]{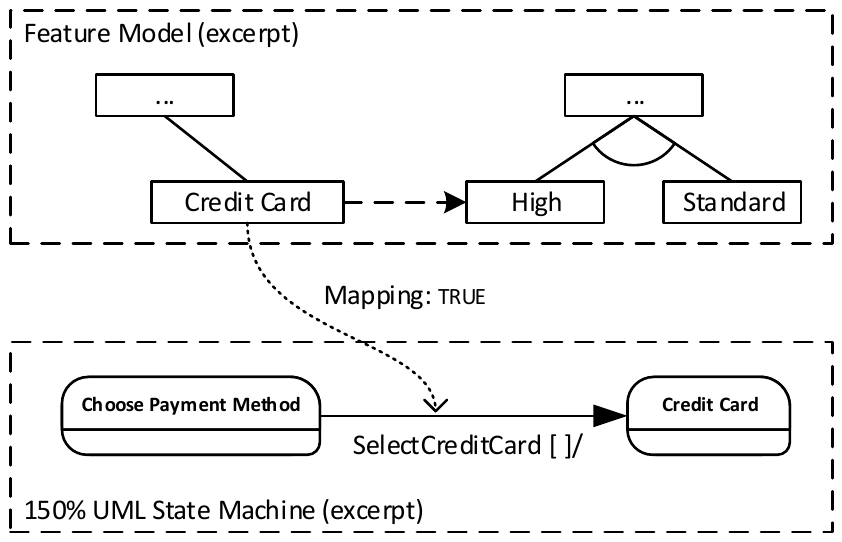}
\captionof{figure}{SPL Design with Negative Variability.}
\label{fig:mapppingmodel}
\end{minipage}%
\end{figure}

A feature model captures the system's variation points in a concise form.
Its elements, however, are only symbols~\cite{Czarnecki.2005}.
Their semantics has to be provided by mapping them to models with semantics.
Such a mapping can be defined using an explicit mapping model.
A mapping model consists of relations from feature model elements to domain model elements. 
We refer to the tuple of feature model, mapping model, and domain model as SPL specification.

Product models or code can be materialized from the SPL specification by providing a configuration.
A configuration assigns a valuation to every feature in the feature model, denoting the presence or absence of the mapped elements.
The valuation must not violate the constraints imposed by the feature model. 

Based on this setup three paradigms have established for specifying SPLs. These paradigms will be briefly introduced as follows.

\subsubsection{Negative Variability}

In this case, the domain model is designed in terms of a so called 150\% model.
A 150\% model contains every element that is used in at least one product configuration and, thus, subsumes every possible product~\cite{Gronniger.2008} (Fig.~\ref{fig:variabilitydelta}a). 
We consider the combination of a feature model, a mapping model, and a UML model as \emph{SPL specification}.

Each mapping in the mapping model maps a single feature to a set of  transitions.
Additionally, each mapping has a Boolean flag that indicates whether the mapped model elements are part of the product when the feature is selected (\emph{true}) or unselected (\emph{false}).
Figure~\ref{fig:mapppingmodel} shows an excerpt of the eShop specification, where parts of the feature model are depicted in the upper half and parts of the state machines payment process are shown in the lower half. 
In between, we find a mapping, denoted by a dotted edge, from feature "Credit Card" to the transition labeled as "SelectCreditCard[]/".

\subsubsection{Positive Variability}

In contrast to negative variability to design the domain model, positive variability starts with a minimal core that contains features that are common to all possible products. From this starting point additional features will be added by a designer (Fig.~\ref{fig:variabilitydelta}b). 

\subsubsection{Delta Modeling}
Designing products in SPL engineering using positive or negative variability is called feature-oriented. 
In contrast to these paradigms, there is another approach which is referred to as delta modeling (also delta-oriented programming)~\cite{Schaefer.2010}. 
Using delta modeling for the purpose of designing SPLs, two parts are needed. The first is a core module, that comprises a set of features that represent a valid product. The second part is a set of delta modules which specify changes that will be applied to the core module. These changes can either be the construction (add) or destruction (remove) of features (Fig.~\ref{fig:variabilitydelta}c).


\subsection{Mutation Analysis}
Mutation analysis (also mutation testing) as introduced by DeMillo et al.~\cite{Demillo.1980} is a error-based testing technique with the intended purpose to assess the quality of tests that will be applied to a system.

The process of mutation analysis seeds errors into software by creating modified versions of the original software, where each created version contains one error. 
After that existing test cases are used to execute the defective versions (\textit{mutants}) with the goal to distinguish the defective ones (\textit{to kill a mutant}) from the original software. 
The ratio of killed mutants to generated mutants is called \textit{mutation score}, that will be computed after the execution of all test cases. 
The main goal of the test designer is to achieve the highest possible mutation score~\cite{Offutt.2001, Jia.2011}. 

Though mutation operators are applied to introduce errors, there is the chance, that the resulting mutant offers the same behavior like the original. 
This type of mutants are referred as \textit{hidden mutants}. 
Although the detection of hidden mutants is an undecidable problem, hidden mutants are supposed to be removed from the mutation analysis before scoring is performed~\cite{Jia.2011}.

According to Jian and Harman~\cite{Jia.2009} we can distinguish multiple kinds of mutants that can be created. 
The simplest ones and already mentioned are \textit{first-order mutants} that have only one introduced error. 
Even if first-order mutants can be killed during the process of mutation testing, this does not guarantee that a combination of two (or even more) mutants will also be detected by the test suite. 
Such combined mutants are referred as \textit{higher-order mutants}.

\subsection{Potential Errors and Mutations}
In mutation analysis, defective software versions are derived from a set of potential errors a human can make during software development. 
Potential errors are implemented as mutation operators, which are applied to the original software for introducing errors.
The mutation operator's design affects the validity of the resulting mutation scores and the costs for testing by means of the amount of mutants to create and the number of tests to execute against them. 
Thus, we apply the following four guiding principles for creating mutation operators~\cite{Black.2000, Woodward.1993}:
\begin{enumerate}
\item Mutation categories should model potential error.
It is important to recognize different types of error. 
In fact, each mutation operator is designed to model errors belonging to the corresponding error class.
\item Only simple, first-order mutants should be generated.
These mutants are produced by making exactly one syntactic change to the original specification. This restriction is justified by the \textit{coupling effect} hypothesis which says that the test sets that detect simple mutants will also detect more complex mutants~\cite{Offutt.1992}.
\item Only syntactically and semantically correct mutants should be generated.
Some mutations may result in an illegal expression, such as division by 0. Such mutants should not be generated.
\item Do not produce too many mutants.
This includes some practical restrictions. For example, do not replace a relational connector with its opposite, if for other mutants a term negation operator is applied, since both mutants are semantically equivalent.
\end{enumerate}

From other mutation systems~\cite{Belli.2011, Aichernig.2013, Fabbri.1994}, we identified the following general categories for model-based mutation operators:
\begin{enumerate}\item Model element deletion: a model designer forgets to add a model element, e.g. a feature, a mapping, or a transition
\item Model element insertion: a model designer inserts a superfluous model element, e.g. a feature, a mapping, or a transition
\item Property change: a model designer chooses a wrong value for a property of a model element, e.g. mandatory feature instead of optional, inverse value for a feature's status, or wrong transition target.
\end{enumerate}
For each model element-type, like mappings, transitions, guards, etc., one can check for applicable categories and implement mutation operators accordingly.

\section{Potential Errors in Model-Based Product Line Engineering}\label{sec:errors}
\label{sec.potentialerrors}

In this contribution, we focus on errors in the feature mapping.
The feature mapping has a major impact on the outcome of the product line's materializations, however the design is complex and error-prone. 
We identify potential errors in a systematic way by checking each modeling paradigm for possibilities to add superfluous or omit necessary elements or change the value of an element's attribute.
For each potential error we discuss its effects onto the materializations.

Furthermore, according to the consequences of each errors for affected products, we assign one of the following four types to each potential error: 
\begin{enumerate}[leftmargin=*,labelindent=20pt, itemsep=0.5mm]
\item[\emph{add}] extends the behavior of affected products.
\item[\emph{omit}] restricts the behavior of affected products.
\item[\emph{alter}] extends and restricts the behavior of some products.
\item[\emph{mix}] extends, restricts, or both the behavior of affected products, depending on the model's contents.
\end{enumerate}

\subsection*{Negative Variability}
In the negative variability paradigm, we identify the following model elements for potential errors from the feature mapping model: mappings, their attribute feature value, mapped feature, and the set of mapped elements.
The errors which can be made on these model elements and their effects are as follows: 
\begin{enumerate}[label=N\arabic*), leftmargin=0pt,itemindent=23pt]
\item \emph{Omitted mapping:} 
a necessary mapping is left out by its entirety. 
Mapped elements will be part of every product unless they are restricted by other features.
As a result, some or all products unrelated to the particular feature will include superfluous behavior. 
Products including the mapped feature are not affected, since the behavior was enabled anyway.

\item \emph{Superfluous mapping:}
a superfluous mapping is added, such that a previously unmapped feature is now mapped to some domain model elements.
This may also include adding a mapping for an already mapped feature, but with inverted feature value.
Adding a mapping with feature value set to \emph{true} results in the removal of elements from products unrelated to the mapped feature.
Contrary, adding a mapping with feature value set to \emph{false} removes elements from any product which the mapped feature is part of.
In any case the behavior of at least some products is reduced.

\item \emph{Omitting a mapped element:} 
a mapped model element is missing from the set of mapped element in a mapping.
Subsequently, a previously mapped element will not only be available in products which the said feature is part of, but also in products unrelated to this feature.
As a result, some products offer more behavior than they should or contain unreachable model elements. 

\item \emph{Superfluously mapped element:}
an element is mapped although it should not be related to the feature it is currently mapped to.
As a result the element becomes unavailable in products which do not include the associated feature. 
The product's behavior is hence reduced.


\item \emph{Swapped feature:}
the associated features of two mappings are mutually exchanged.
Subsequently, behavior is exchanged among the two features and thus, affected products offer different behavior than expected.
The result is the same as exchanging all mapped elements among two mappings.

\item \emph{Inverted feature status:}
the bit-value of the feature value attribute is flipped.
The mapped elements of the affected mapping become available to products where they should not be available.
At the same time, the elements become unavailable in products where they should be.
For example, if the feature value is true and is switched to false, the elements become unavailable to products with the associated feature and available to any product not including the said feature.
Of course, other feature mappings to the same element(s) must still be considered.

\end{enumerate}

\subsection*{Positive Variability}
In SPL modeling with positive variability, a mapping is a bijection between features and modules composed from domain elements. 
Potential errors in the feature mapping models can be made at: mappings, mapped feature, and mapped module.
We identify the following potential errors:

\begin{enumerate}[label=P\arabic*), leftmargin=0pt,itemindent=21pt]
\item \emph{Omitted mapping:} 
a necessary mapping is missing in its entirety.
This appears to us to be an unrealistic scenario, since one can automatically check for all modules being mapped to some feature.
But if we consider the case of a missing mapping, products with the associated feature would be missing the modules functionality.

\item \emph{Superfluous mapping:}
a superfluous mapping is added. 
Similar to the above, this is an unrealistic scenario for same reason: all modules should be mapped exactly once.
In a model-based environment, this check should be easily automatable.
However, if adding a superfluous mapping is possible, more behavior becomes enabled in products containing the mapping's feature.

\item \emph{Swapped modules:}
the associated modules of two mappings are mutually exchanged.
As a result, all products containing one of the two features, but not the other, offer not the expected behavior.
In contrast, all products containing none or both features behave as expected.

\item \emph{Swapped features: }
the associated features of two mappings are mutually exchanged.
The result is the same as above for swapped modules.


\end{enumerate}

\subsection*{Delta Modeling}
For other paradigms, like delta-modeling~\cite{Schaefer.2010}, we make similar observations. In contrast to positive variability models, delta-oriented variability models start from an actual core product, instead of a base module. 
From this on, only the differences from one product to another are defined by \emph{deltas}. In delta-modeling, mapping multiple features to the same delta is allowed. 
A delta may add elements to and remove elements from the core product at the same time.
As potential points of errors in delta-modeling we identify deltas, a delta's set of mapped features, its set of removed elements from the base product, and its set of added elements.

\begin{enumerate}[label=D\arabic*), leftmargin=0pt,itemindent=23pt]
\item \emph{Omitted delta:}
the product line model misses an entire delta definition.
Products containing features of the missing delta may lack behavior or offer to much of it.
This depends on whether the delta removes and/or adds elements to the base product.

\item \emph{Superfluous delta:}
an unnecessary delta is added. 
As a result, products containing the associated feature(s) will offer additional behavior.
Also, affected products might lack behavior if the delta removes elements.

\item \emph{Omitted feature:}
a necessary feature from the set of mapped features is missing.
If no feature is left, the delta is not mapped at all which can be statically verified. 
If otherwise the set still contains at least one feature, any product containing the current set of mapped features but not the missing feature, offers too much or too few behavior.
In some cases, the set of mapped features and the affected elements may collide with another delta, which is again statically verifiable.

\item \emph{Superfluous feature:}
an additional feature is added to a delta's already complete set of mapped features.
As a result, the delta will be available in less products.
If the added feature mutually excludes one of the already mapped features, the delta will be applicable to no product at all.
A static check can be used to validate that a set of features is satisfiable by some product.
Only products containing the correct set of features but not the superfluous, are affected by this error.
Affected products may offer more or less behavior.


\item \emph{Omitted base element:}
a delta's set of base elements is missing an element.
In consequence, too few elements are removed from the core product by this delta to match the product's specification.
Thus any product containing the features from this delta offers too much behavior.

\item \emph{Superfluous base element:}
a delta's set of base elements contains additional elements.
This will remove more elements than necessary from the products affected by this delta.
Hence, these products offer too few behavior.


\item \emph{Omitted delta element:}
an element from the set of delta elements in a delta is missing.
As a result, any product containing the delta's features offers less behavior than specified.

\item \emph{Superfluous delta element:}
an element from a delta's set of delta elements is missing.
In consequence, the products containing the delta's features offer more behavior than specified.



%
\end{enumerate}

\section{Product Line Test Assessment}
\label{sec.productlinetestassessment}
As laid out in section~\ref{sec:errors}, other errors can be made in model-based SPL engineering than in contrast to non-variable systems engineering.
Furthermore, current test design methods and coverage criteria are not prepared to deal with these errors.
To show the validity of our argument, we propose a mutation system for SPLs.
It is specifically designed to assess test quality, by means of error detection capability (EDC), for the whole product line rather than for single systems.
But mutation systems for SPLs need novel mutation operators.
The reason for this is the separation of concerns in SPL engineering, where variability and domain engineering are split into different phases and models.

Mutation operators defined for non-variant systems cannot infer mutants including modules from other products, since this information is only available during domain engineering.
However, we expect a high-quality test suite to detect such errors. 
Hence, we also propose new mutation operators based on the potential errors, we identified in section~\ref{sec:errors}.
For conciseness, we only consider potential errors from negative variability modeling for implementation.

\subsection{Mutation System for SPLs}
Performing mutation analysis on SPL tests is different from non-variant system tests, since in contrast to conventional mutation systems, a mutated SPL specification is not executable per se.
Thus, testing cannot be performed until a decision is made towards a set of products for testing.
This decision depends on the SPL test suite itself, since each test is applicable to just a subset of products.

\begin{figure}
\centering
\includegraphics[width=\columnwidth]{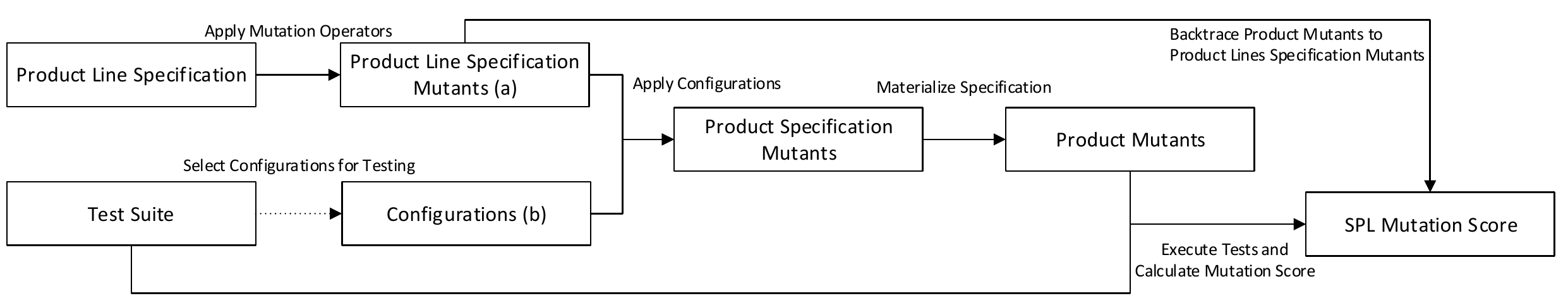}
\caption{\label{fig:mutationprocess}Mutation Process for SPLs}
\end{figure}

In Figure~\ref{fig:mutationprocess}, we depict a mutation process for assessing SPL test suites, which addresses this issue.
Independently from each other, we gain (a) a set of SPL specification mutants by applying mutation operators to the SPL specification and identify (b) a set of configurations describing the applicable products for testing.
We apply every configuration in (b) to every mutant in (a), which returns a new set of product specification mutants.
Any mutant structurally equivalent to the original product specification is immediately removed and does not participate in the scoring.
The specification mutants are easily materialized into product mutants and finally, tests are executed.
Our mutation scores are based on the SPL specifcation mutants, hence we established bidirectional traceability from any SPL specification mutant to all its associated product mutants and back again.
If a product mutant is killed by a test, we backtrack its original SPL specification mutant and flag it as killed.
The final mutation score is then calculated from the killed and the overall number of SPL specification mutants.

\subsection{SPL Mutation Operators}
Here, we present mutation operators for feature mapping models with negative variability.
Furthermore, we enrich the mutation system by standard state machine operators and apply them on domain-level as well.
For each operator, we describe how it was identified and its notion. 
Also, we discuss potentially invalid and hidden mutants resulting from each operator.

\subsubsection{Feature Mapping}
We design the mutation operators according to the potential errors identified in section~\ref{sec:errors}.
We do not consider inserting superfluous mappings as in this case it remains unclear which and how many UML elements should be selected for the mapping.
We assume that this, if not carefully crafted, will lead to mostly invalid mutants.

\begin{description}[style=unboxed,leftmargin=0cm]
\item[\textnormal{\emph{Delete Mapping (DMP)}}]
The deletion of a mapping will permanently enable the mapped elements, if they are not associated to other features that constrain their enabledness otherwise. 
In our examples, no invalid mutants were created.
However, for product lines that make heavy use of mutual exclusion (Xor and excludes) this does not apply.
The reason for this are competing UML elements like transitions that would otherwise never be part of the same product.
Multiple enabled and otherwise excluding transitions are possibly introducing non-determinism or at least unexpected behavior.

Some products mutants created with this operator might behave equivalent to an original product. 
This is the case for all products that include the feature for which the mapping was deleted.

\item[\textnormal{\emph{Delete Mapped Element (DME)}}]
This operator deletes a mapped UML element from a mapping in the feature mapping model.
It resembles the case, where a modeler forgot to map a UML element that should have been mapped.

Similar to the delete mapping operator, this operator may yield non-deterministic models, where otherwise excluding transitions are concurrently enabled.
Product mutants equivalent to the original product model can be derived, if the feature associated to the deleted UML reference is part of the product.

\item[\textnormal{\emph{Insert Mapped Element (IME)}}]
This operator inserts a new UML element to the mapping. 
This is the contrary case to the operators defined before, where mappings and UML elements were removed.
However, inserting additional elements is more difficult than deleting them, since a heuristic must be provided for creating such an additional element.
We decided to copy the first UML element reference from the subsequent mapping.
If there are no more mappings, we take the first mapping.
This operator is not applicable if there is just one mapping in the feature mapping model.

Again, there is a chance of creating invalid mutants:
If a UML element reference is copied from a mutually excluded mapping, the resulting model may be invalid due to non-determinism.

\item[\textnormal{\emph{Swap Feature (SWP)}}]
Swapping features exchanges the mapped behavior among each other. 
This operator substitutes a mapping's feature by the following mapping's feature and vice versa.
The last feature to swap is exchanged with the very first of the model.

Non-deterministic behavior and thus invalid models may be designed by this operator.
This is due to the fact that the mutation operator may exchange a feature from a group of mutually exclusive features by an unrestricted feature.
In consequence, the previously restricted feature is now independent, while the unrestricted feature joins the mutually exclusive group.
This may concurrently enable transitions which results in non-deterministic behavior.

\item[\textnormal{\emph{Change Feature Value (CFV)}}]
This operator flips the feature value of a mapping.
A modeler may have selected the wrong value for this boolean property of each mapping.

The operator must not be applied to a mapping, if there is a second mapping with the same feature, but different feature value.
Otherwise, there will be two mappings for the same feature with the same feature value, which is not allowed for our feature mapping models.

This operator may yield invalid mutants, if it is applied to a mapping that excludes another feature. 
In that case, two otherwise excluding UML elements can be present at the same time, which may result in invalid models, e.g. two default values assigned to a single variable or concurrently enabled transitions.


\end{description}

\subsubsection{UML State Machine}

In the past 20~years, many mutation operators for transition-based systems were defined~\cite{Fabbri.1994, Offutt.2003, Belli.2006, Belli.2008}. 
Here, we limit ourselves to the design of operators based on transitions as these may have the strongest impact on the behavior of the SUT.
We do not design operators that can be mimicked by the combination of two of them. 
In particular, we do not consider the exchange of an element by another, since this can easily be mimicked by removing and inserting the removed element at another point in the model.

We identified five targets for mutation: (i) remove the entire transition, change its (ii) target state, as well as mutating its (iii) triggers, (iv) guard, and (v) effect.
The latter three can be mutated according to the three defined categories delete, add and change. 
Though in this contribution omitted the category change for simplicity. 

%

For all mutants created by the here presented operators, there is a chance of materializing mutants behaving equivalent to the original product.
This is the case, when the mutated element is part of disabled feature.
Of course, hidden mutants -- if detected -- will be excluded from the scoring.

In general, we will not apply any class mutation to our UML state machines~\cite{Kim.2000}.
The system's logic is designed in the state machine diagrams, while the classes are merely containers for variables and diagrams.

\begin{description}[style=unboxed,leftmargin=0cm]
\item[\textnormal{\emph{Delete Transition (DTR)}}]
Deletes a transition from a region in a UML state machine. 
This operator might create invalid UML models, if not enough transitions remain on a pseudo-state (fork, join, junction, and choice)~\cite[p.555]{ObjectManagementGroup(OMG).2011}.

\item[\textnormal{\emph{Change Transition Target (CTT)}}]
Changes the target of a transition to another state of the target state's region.
This operator is only applicable if the region has more than one state.

\item[\textnormal{\emph{Delete Effect (DEF)}}]
Deletes the entire effect from a transition.
We consider sending signals to the environment or other components to be part of a transition's effect, 
hence they are deleted as well.

\item[\textnormal{\emph{Delete Trigger (DTI)}}]
Deletes a transition's trigger.
Only a single trigger is deleted at a time, but every trigger is deleted once.

\item[\textnormal{\emph{Insert Trigger (ITG)}}]
Copies an additional trigger to a transition.
The trigger is copied from another transition within the same region.
This may lead to non-deterministic behavior if both transitions, the source transition of the trigger and the mutated transition, are outgoing transitions of the same state.

\item[\textnormal{\emph{Delete Guard (DGD)}}]
Deletes the entire guard of a transition. 
This may lead to non-deterministic behavior of the state machine, if another transition is enabled simultaneously.
Furthermore, in the case of transitions without triggers and where source and target are the same state, this operator leads to infinite looping of the state machine over the mutated transition. 
The reason for this behavior is UML's run-to-completion semantic, where an enabled transition without triggers is immediately traversed.

\item[\textnormal{\emph{Change Guard (CGD)}}]
Changes a guard's term by exchanging operators or substituting boolean literals by their inverse.
Our CGD operator supports 30 different arithmetic, relational, bitwise, compound assignment, and logic operators.
Furthermore, literal "null" is exchanged by "this".

This may cause mutants with non-deterministic behavior, whenever two transition become concurrently enabled due to the manipulation of one of their guards.

\end{description}

\subsection{Evaluation}
We created three example product lines for performing a mutation analysis on them.
We designed the test suite for each example automatically by applying model-based testing techniques. 
In particular, we used product line-centered test design (PLC) from our SPLTestbench as defined in~\cite{Lackner.2014}, where tests are designed from the SPL specification.
In contrast to product-centered test approaches, where tests are designed from selected product specifications, the PLC approach selects products for testing after the test design phase.
This improves coverage of the state machine, since the coverage criteria are applied onto the whole SPL specification.

We chose all-transitions coverage for selecting the tests.
A test generator then automatically designed the tests and outputs XML-documents.
From the tests, SPLTestbench selected variants for testing and materialized them from the mutated SPL specifications into product specification mutants.

Since our examples lack implementations, we decided to generate code from the product specification mutants and run the tests against them.
Therefore, we developed and employed a code generator for transforming individual product specifications into Java.
Another transformator generates executable JUnit code from the tests which we gained from the test generator.
The mutation systems then collects all the code artifacts, executes the tests against the product code, and finally reports the mutation scores for all tests and for every operator individually.
All of the transformations above and the mutation system are part of our SPLTestbench.

Generating code \emph{and} tests from the same basis for testing the code is not feasible in productive environments, since errors propagate from the basis to code \emph{and} tests.
However in our case, tests are executed against code derived from mutated artifacts, which are different from the original.

\subsubsection{Examples}\label{subsec.casestudies}
Our examples represent three kinds of systems: 
an e-commerce shop (eShop), which makes heavy use of signals but with only few guards, a ticket machine (TicketMach) that uses less signals and in contrast more guards, and lastly, an alarm system (AlarmSys), which uses various signals and guards and is more variant-rich than the other two case studies.

The eShop is a fictional example designed by ourselves, which is comprised of 10 features offering 20 different variants.
A customer can browse the catalog of items, or if provided, use the search function.
Once the customer put items into the cart, he can checkout and may choose from up to three different payment options, depending on the eShop's configuration.
The transactions are secured by either a standard or high security server.
A constraint ensures that credit card payment is only offered if the eShop also implements a high security server.

The TicketMach example is adopted from Cichos et al.~\cite{Cichos.2012}.
The functionality is as follows: a customer may select tickets, pay for them, receive the tickets, and collect change.
The feature model has a root feature with three sub-features attached to it; all of them are optional with no further constraints, thus it offers eight variants.
Depending on the selected features, the machine offers reduced tickets, accepts not only coins but also bills, and/or will dispense change. 

From Cichos et al.~\cite{Cichos.2011} we also adopted and extended the AlarmSys example.
Currently, it consists of 12 features and offers 42 variants.
The alarm may be set off manually or automatically by a vibration detector. 
Both features are part of an or-group and, thus, at least one of the two features must be present in every product.
In the event of an alarm, a siren or a warning light will indicate the security breach. 
When the vibration does not stop after a predefined period of time, the system optionally escalates the alarm by calling police authorities and/or sending photos of evidence. 
Additionally to its alarming functionality, the  AlarmSys SPL provides a feature for taking a photo of any operator that configures the system for security measures.

\subsubsection{Results}
We were able to assess the test quality for all three test suites derived from the examples. 
Here, we present our results.
For each mutation operator we measured the amount of detected mutants based on the SPL specification.
In addition, we assessed accumulated mutation scores for each example over all mutation operators and vice versa, the accumulated results for each mutation operator over all examples.
The detailed results for feature mapping operators can be read from Table~\ref{mappingscore} and for UML operators from Table~\ref{umlscore}.

\newlength{\oneDec}
\setlength{\oneDec}{3.5pt}
\begin{table}
\centering
    \parbox[t]{.45\linewidth}{
    \centering
\caption{Mapping Operator Scores per Mutation Operator in~\% and Accumlated Scores (Acc)}
\label{mappingscore}
\scalebox{.78}{
    \begin{tabular}{crrrr} \toprule
Op. & \multicolumn{1}{c}{eShop} & \multicolumn{1}{c}{TicketMach} & \multicolumn{1}{c}{AlarmSys} & \multicolumn{1}{c}{Acc}\\ \midrule
DMP & 0.00\hspace{\oneDec} (4) 		& 0.00 (5) 	& 0.00 \hspace{\oneDec}(8) 		& 0.00\\ 
DME & 0.00 (14) 					& 0.00 (8) 	& 0.00 (21)						& 0.00\\ 
IME & 75.00\hspace{\oneDec} (4) 	& 40.00 (5) & 50.00 \hspace{\oneDec}(8) 	& 52.94\\
SWP & 100.00\hspace{\oneDec} (4) 	& 60.00 (5) & 62.5 \hspace{\oneDec}(8) 		& 70.59\\ 
CFV & 100.00\hspace{\oneDec} (4) 	& 100.00 (5) & 87.50 \hspace{\oneDec}(8) 	& 94.12\\ 
\midrule
Acc & 36.67 (30) 					& 35.71 (28) & 30.19 (53) 			& 33.33\\ 
\bottomrule
\end{tabular}}
}
\hfill
    \parbox[t]{.45\linewidth}{
\centering
\caption{UML Operator Scores per Mutation Operator in~\% and Accumlated Scores (Acc)}\label{umlscore}
\scalebox{.78}{
\begin{tabular}{crrrr} \toprule
Op. & \multicolumn{1}{c}{eShop} & \multicolumn{1}{c}{TicketMach} & \multicolumn{1}{c}{AlarmSys} & \multicolumn{1}{c}{Acc}\\ \midrule
DTR & 89.29 (28) 			& 84.21 (19) 	& 63.16 (19) 				& 80.30\\ 
CTT & 64.29 (28)			& 63.16 (19) 	& 36.84 (19) 				& 56.06\\ 
DEF & 100.00 (16) 			& 82.35 (17) 	& 61.54 (13) 				& 82.61\\
DTI & 82.61 (23) 			& 100.00 (13) 	& 94.12 (17) 				& 90.57\\ 
ITG & 20.83 (24) 			& 27.78 (18) 	& 16.67 (18) 				& 21.67\\ 
DGD & 0.00\hspace{\oneDec} (1) & 42.86 (14) 	& 50.00\hspace{\oneDec} (2) 	& 41.18\\ 
CGD & 100.00\hspace{\oneDec} (2) & 68.75 (48)	& 90.00 (10) 	& 73.33\\ 
\midrule
Acc & 69.67 (122) & 66.89 (148)	& 57.17 (98) 	& 65.21\\ 
\bottomrule
\end{tabular}
}}

\end{table}

Furthermore, we tracked for every example the number of original products selected for testing, generated product line mutants, and materialized product mutants.
Test-wise we counted tests, test steps by means of stimuli and expected reactions in all tests, tests executed against all product mutants, and the number of failed tests during test execution.

For the eShop example, SPLTestbench selected four products for testing. 
Independent from this, the mutation system generated 30 product line mutants and 96 product mutants for the mapping mutation operators. For the state machine mutation operators it generated 122 product line mutants and 478 product mutants.
Every test from the 13 tests for this example were executed against every suitable mutant.
This results in 302 test executions for the mutants created by the mapping mutation operator and 1553 test execution for state machine mutation operators.
Ultimately, 20 tests for mapping operators and 283 tests for state machine operators failed, killing 69.67\% and 36.67\% of the mutants.

Analog to the eShop, we executed less tests and generated less product mutants for the feature mapping operators: 252 tests were executed against 56 product mutants.
The tests yield an even lower mutation score of 35.71\% than for the eShop case study.

In case of the AlarmSys, we executed 537 tests against 278 product mutants created by the mapping mutation operators and 1168 tests against 585 product mutants created by the state machine mutation operators.
Eventually, 37 and 123 tests failed, killing 30.19\% and 57.17\% of the mutants, respectively. The results are summarized in Table~\ref{summaryMapping} and~\ref{summaryUML}.

\begin{table}    
\parbox[t]{.45\linewidth}{
\centering
\caption{Summarized Results for Mapping  Operators}\label{summaryMapping}
\scalebox{.76}{
\begin{tabular}{lrrr} \toprule
\multicolumn{1}{c}{} & \multicolumn{1}{c}{eShop} & \multicolumn{1}{c}{TicketMach} & \multicolumn{1}{c}{AlarmSys}\\ \midrule
Products for testing	& 4 & 4 & 6\\ 
Product line mutants	& 30 & 28 & 53\\
Product mutants 		& 96 & 56 & 278 \\
Tests 			& 13 & 9 & 12\\
Test steps 		& 103 & 68 & 62\\
Tests executed 	& 302 & 252 & 537\\ 
Failed Tests 	& 20 & 30 & 37\\ 
\bottomrule
\end{tabular}
}}
\hfill
    \parbox[t]{.45\linewidth}{
\centering
\caption{Mutation Results for State Machine  Operators}\label{summaryUML}
\scalebox{.76}{
\begin{tabular}{lrrr} \toprule
\multicolumn{1}{c}{} & \multicolumn{1}{c}{eShop} & \multicolumn{1}{c}{TicketMach} & \multicolumn{1}{c}{AlarmSys}\\ \midrule
Products for testing	& 4 & 4 & 6\\ 
Product line mutants	& 122 & 148 & 98\\
Product mutants 		& 478 & 296 & 585 \\
Tests 					& 13 & 9 & 12\\
Test steps 				& 103 & 68 & 62\\
Tests executed 			& 1553 & 1332 & 1168\\ 
Failed Tests 			& 283 & 272 & 123\\ 
\bottomrule
\end{tabular}
}}
\end{table}

\section{Related Work}
\label{sec.relatedwork}

Mutation analysis for SPLs seems to be a rather new topic.
To our knowledge, there is no publication dealing with mutation operators on all model artifacts of a SPL specification.
Though, Henard et al. proposed two mutation operators for feature models based on propositional formulas in~\cite{Henard.2013}.
They employ their mutation system for showing the effectiveness of dissimilar tests, in contrast to similar tests.
For calculating dissimilarity, the authors provide a distance metric to evaluate the degree of similarity between two given products.

In contrast, mutation analysis for behavioral system specifications, e.g. finite state machines, is established since two decades.
Fabbri et al. introduced mutation operators for finite state machines in~\cite{Fabbri.1994}.
In addition to our operators, they also consider adding states and the exchange of elements (event, guard, effect) by another.
Belli and Hollmann provide mutation operators for multiple formalism: directed graphs, event sequence graphs~\cite{Belli.2006}, finite state machines~\cite{JeffOffutt.2003}, and basic state charts~\cite{Belli.2008}.
They conclude, that there are two basic operations from which most operations can be derived: omission and insertion.
Also for timed automata, mutation operators can be found in~\cite{Aichernig.2013}.

In~\cite{Stephenson.2004} Stephenson et al. propose the use of mutation testing for prioritizing test cases from a test suite in a SPL environment.
Unfortunately, the authors provide no evaluation of their approach.

\section{Conclusions}
\label{sec.conclusions}
In this contribution, we lifted mutation analysis to the product line level.
We defined and investigated mutation operators for feature models, mapping models, and UML models.
As opposed to product-based mutation analysis, our mutation operators are based on the SPL specification.
This allows us to mimic realistic errors made by humans during modeling a SPL.
To our knowledge, this is the first step towards a qualitative evaluation of SPL tests, which is based on the SPL's specification.

Our results for the three examples are as expected for most of the mutation operators.
As predicted, mutation operators contributing superfluous behavior are hard to detect for conformance tests.
Such mutations are DMP (0\%) and DME (0\%) on feature mappings and ITG (21.67\%) on domain models.
For most of the other operators we gain scores above 70\%, which is in the expected range for all-transitions coverage~\cite{Weileder.2009}.
For DGD and CTT mutations the tests score surprisingly low results.
Here, further investigations seem necessary.

In conclusion, we identified a lack of error detection capability in standard test procedures for SPLs.
Even simple errors are not detectable, neither by all-transitions, MC/DC as for safety-critical system, nor any other conformance test procedure. 
As indicated, the results are applicable to at least the here surveyed SPL engineering paradigms negative/positve variability and delta modeling.
We assume, other paradigms suffer from this lack as well.
Unfortunately, current procedures for negative testing, which could potentially detect such errors are still not enabled for SPLs.
Thus, future work will proceed to enable negative testing procedures for SPLs. 

In~\cite{Lackner.2014}, we described product-centered and product line-centered test design processes for SPLs. 
We plan to employ this mutation system for assessing the quality of the test suites generated by the different test design methods. 
From the results we hope to gain general directions towards favorable test design methods and processes by means of error detection capability, test effort, and efficiency.

Furthermore, we want to investigate higher-order mutation operators, that combine more than one change at a time to a product. 
For this purpose we need co-adaptations, so that the parts that constitute the SPL, here feature model, mapping/delta model, and domain model, are adapted to preserve consistency when one of the parts changes.
For example, such an adaptation is necessary after the deletion of a feature to ensure that there is no broken feature reference in related mappings. 
In~\cite{Schmidt.2013}, we presented a prototype for co-adaptations in another model-based scenario, namely domain-specific language development.

\section*{Acknowledgments}
This work is supported by grants from Deutsche For\-schungs\-ge\-mein\-schaft, Graduiertenkolleg METRIK (GRK 1324).

\bibliographystyle{eptcs}
\bibliography{biblio}
\end{document}